\documentclass[prb,aps,epsf,twocolumn,showpacs,superscriptaddress]{revtex4}
\usepackage{graphicx}
\usepackage{epsfig}
\usepackage{latexsym}
\usepackage{amsmath}
\begin{document}
\newcommand{\beq}{\begin{equation}}
\newcommand{\beqr}{\begin{eqnarray}}
\newcommand{\eeqr}{\end{eqnarray}}
\newcommand{\eeq}{\end{equation}}
\newcommand{\s}{{\sigma}}
\newcommand{\e}{{\varepsilon}}
\newcommand{\om}{{\omega}}
\newcommand{\Om}{{\Omega}}
\newcommand{\D}{{\Delta}}
\newcommand{\de}{{\delta}}
\newcommand{\al}{{\alpha}}
\newcommand{\ga}{{\gamma}}
\newcommand{\La}{{\Lambda}}
\newcommand{\be}{{\beta}}
\newcommand{\psib}{{\bar{\psi}}}
\newcommand{\rb}{{\bar{\rho}}}
\newcommand{\phib}{{\bar{\phi}}}
\newcommand{\dt}{{\Delta}}
\newcommand{\dva}{{\frac{\vp\times\va}{2\pi}-\rb}}
\newcommand{\dvab}{{\frac{\vp\times(\va-\vb)}{4p\pi}}}
\newcommand{\w}{{\omega}}
\newcommand{\zh}{{\hat{z}}}
\newcommand{\qh}{{\hat{q}}}
\newcommand{\vA}{{\vec{A}}}
\newcommand{\va}{{\vec{a}}}
\newcommand{\vrr}{{\vec{r}}}
\newcommand{\vj}{{\vec{j}}}
\newcommand{\vE}{{\vec{E}}}
\newcommand{\vB}{{\vec{B}}}
\def\bA{{\mathbf A}}
\def\bm{{\mathbf m}}
\def\bsig{{\mathbf \sigma}}
\newcommand{\etab}{\mbox{\boldmath $\eta $}}
\def\bB{{\mathbf B}}
\def\bp{{\mathbf p}}
\def\bI{{\mathbf I}}
\def\bn{{\mathbf n}}
\def\bM{{\mathbf M}}
\def\bG{{\mathbf G}}
\def\bq{{\mathbf q}}
\def\bp{{\mathbf p}}
\def\br{{\mathbf r}}
\def\bx{{\mathbf x}}
\def\bs{{\mathbf s}}
\def\bS{{\mathbf S}}
\def\bQ{{\mathbf Q}}
\def\bq{{\mathbf q}}
\def\bs{{\mathbf s}}
\def\bi{{\mathbf i}}
\def\bj{{\mathbf j}}
\def\bB{{\mathbf B}}
\def\bl{{\mathbf l}}
\def\bPi{{\mathbf \Pi}}
\def\bJ{{\mathbf J}}
\def\bR{{\mathbf R}}
\def\bz{{\mathbf z}}
\def\ba{{\mathbf a}}
\def\bk{{\mathbf k}}
\def\bK{{\mathbf K}}
\def\bP{{\mathbf P}}
\def\bg{{\mathbf g}}
\def\bX{{\mathbf X}}
\newcommand{\Psib}{\mbox{\boldmath $\Psi $}}
\newcommand{\thetab}{\mbox{\boldmath $\theta $}}
\newcommand{\sigmab}{\mbox{\boldmath $\sigma $}}
\newcommand{\gammab}{\mbox{\boldmath $\gamma $}}
\newcommand{\vx}{{\vec{x}}}
\newcommand{\vq}{{\vec{q}}}
\newcommand{\vQ}{{\vec{Q}}}
\newcommand{\vd}{{\vec{d}}}
\newcommand{\vb}{{\vec{b}}}
\newcommand{\vp}{{\vec{\partial}}}
\newcommand{\p}{{\partial}}
\newcommand{\gr}{{\nabla}}
\newcommand{\ra}{{\rightarrow}}
\def\dd{d^{\dagger}}
\def\half{{1\over2}}
\def\third{{1\over3}}
\def\twof{{2\over5}}
\def\threes{{3\over7}}
\def\rhob{{\bar \rho}}
\def\ua{\uparrow}
\def\da{\downarrow}
\def\eqa{\begin{eqnarray}}
\def\eea{\end{eqnarray}}
\def\jetp{{\it Sov. Phys. JETP\ }}
\parindent=4mm
\addtolength{\textheight}{0.9truecm}
\title{Composite fermions for fractionally filled Chern bands \\
}
\author{Ganpathy Murthy }
\affiliation{  Department of Physics and
Astronomy, University of Kentucky, Lexington KY 40506-0055   }
\author{R. Shankar}
\affiliation{Department of Physics, Yale University, New Haven CT 06520}

\date{\today}

\begin{abstract}

We address  the question of
whether fractionally filled bands with a nontrivial Chern index in zero external field  could also exhibit a Fractional Quantum Hall Effect
(FQHE). Numerical works suggest this is 
possible. Analytic treatments are complicated by a
non-vanishing band dispersion and a non-constant Berry flux. We
propose embedding the Chern band in  an auxiliary lowest Landau
level (LLL) and then using composite fermions. We find some states which
have no analogue in the continuum, and dependent  on the
interplay between interactions and the lattice. The approach extends to two-dimensional time-reversal invariant topological
insulators.

\end{abstract}

\maketitle

Models with no net magnetic flux but with a quantized Hall conductance
$\sigma_{xy}$ have been known for some time\cite{haldane,volovik}. The
breaking of time-reversal symmetry necessary for $\sigma_{xy}\ne0$
manifests itself as a nontrivial Berry flux for the band in the
Brillouin zone (BZ), which integrates to a nonzero
integer Chern index, leading to a quantized Hall conductance when the
band is full\cite{TKNN}.

Could these "Chern bands" (CBs)\cite{qiwu} exhibit the FQHE at partial
filling in the presence of suitable interactions?  A necessary
condition is a hierarchy of scales where the bandgaps $\Delta$, the
interaction strength $U$, and the bandwidth of the CB $W$ satisfy $\Delta\gg
U\gg W$. Efforts have concentrated on ``flattening'' the
CB\cite{tang,sun,neupert}, where numerical work suggests that FQH-like
states are realized\cite{sun,neupert,sheng,bern,sheng-wang}.  On the
analytical front, Qi \cite{qi} has constructed a basis in which known
FQHE wavefunctions can be transcribed into the CB.

Recently Parameswaran {\em et al} \cite{sid} examined the algebra of
$\bar{\rho}_{CB}(\bq)$, the density operators projected into the
CB. Unlike in the LLL, where the projected density operators obey the
magnetic translation algebra\cite{GMP}, in a CB the algebra does not
close because the Berry curvature is not a constant. Equivalently,
consider $R_{\mu}^{CB}$, the projected position vector of the electron
with matrix elements
\begin{equation}
\big(R_{\mu}^{CB}\big)_{\bk\bk'} = \big(i\frac{\partial} {\partial
  k_{\mu}} + A_{\mu}(\bk)\big)\delta^2(\bk-\bk') 
\end{equation}
where  $A_{\mu}=-i\langle u (\bk) |\partial_{\mu}|u(\bk )\rangle
$ and 
$
\left[R_{\mu}^{CB},R_{\nu}^{CB}\right]_{\bk\bk'} = i{\cal B}_{\mu\nu}(\bk)\delta^2(\bk-\bk')
$
and  the curvature ${\cal B}(\bk)$ is related to the Chern index by
$
C= \frac{1}{2 \pi} \int _{BZ}{\cal B}
$. 

In an LLL problem in the continuum ${\cal B} (\bk )$ is a constant and
the projected electronic guiding center coordinates $\bR^{e}$) have
$c$-number commutation relations
$\left[R_{\mu}^{e},R_{\nu}^{e}\right]=-il^2$, where $l$ is the
magnetic length. This c-number commutator implies the closure of the
magnetic translation algebra\cite{GMP}, the analyticity of
wavefunctions, and the flux attachment that underlies the Laughlin
wavefunction\cite{laugh}, Composite Bosons\cite{zhk}, and Composite
Fermions (CFs)\cite{jain}.

Our main result is that one can use all the beautiful properties of
the LLL, including flux-attachment, CFs, etc., by embedding the CB in
an auxiliary LLL with a periodic potential. The price to be paid is
either a more complicated density operator or the presence of extra
states, but the profit is in the fact that one can carry out dynamical
calculations for an arbitrarily varying ${\cal B}(\bk)$. The
applicability of CFs extends to analogues of the $\nu=\half$ CF-Fermi
Liquid\cite{hlr} or the $\nu=\frac{5}{2}$ state\cite{moore-read} in
CBs. New states for fractionally filled CBs, with no counterparts in
the continuum, are clearly exhibited in this embedding, which comes in
two variants (i) and (ii).

{\underline{\bf Embedding (i)}}: This maps the CB into a modified LLL,
and seems to display the phenomenology of the gapped states seen in
numerics\cite{sun,neupert,sheng,bern,sheng-wang} on fractionally
filled CBs. It also makes an explicit connection to work on fractional
quantum Hall states on a lattice {\it in an external
  field}\cite{kol-read,sorensen,moller}. Consider a set of
noninteracting Landau levels (LLs) with a periodic potential (with the
lattice symmetries of the target CB), with one quantum of electronic
flux per unit cell.  For a periodic potential $V$ on the same scale as
the cyclotron energy $\omega_c$, LL-mixing will induce a set of
dispersing bands with nonuniform Chern density in the BZ. We assume
that by manipulating the harmonics of $V$ we can flatten the energy
band and approximate the Chern density of the CB in the lowest band of
our mixed LLs, which is the modified LLL, or MLLL with bandwidth $W$.

Now turn on repulsive interactions $U\gg W$, and $U\ll
\Delta=max(V,\omega_c)$ and fractionally fill the MLLL, and attach two
units of flux {\it a la} Jain\cite{jain}. In general the CFs will see
$\frac{p'}{q'}$ units of effective flux per unit cell, and each
CF-Landau level (CFLL) will break up into $p'$
subbands\cite{TKNN}. Filling an integer number $n'$ of these subbands
will yield a gapped state. If the $n'=r'p'$ is an integer multiple of
$p'$, these gapped states are adiabatically connected to the principal
fractions $\nu^e=\frac{r'}{2r'+1}$ seen in the LLL\cite{jain}. For
such states, the Hall conductance is identical to the filling factor,
and the lattice seems to be an unimportant detail. On the other hand,
if $n'$ is not an integer multiple of $p'$, the resulting gapped
states rely crucially on the interplay of interactions and the lattice
potential. Such states will not have any analogues in the
translationally invariant LLL, and can display a Hall conductance
which is different from the filling factor\cite{kol-read}.

{\underline{\bf Embedding (ii)}}: This maps the CB into a subband of
the LLL with a periodic potential $V$ with a rational flux $p/q$ per
unit cell, with $p>1$. No other LLs are present.  The LLL breaks up
into $p$ subbands each with degeneracy $1/p$ of the LLL, each
characterized by a Chern index\cite{TKNN}. Identify a subband with
Chern index $1$, which will play the role of the CB in the larger
space. We assume that by manipulating $V$ one can flatten the target
subband so that $W\ll V$, as well as achieve the required flux density
${\cal B}(\bk)$ in the BZ.

Now turn on repulsive interactions $U$ which satisfy $W\ll U\ll V$. We
argue by adiabatic continuity that the CFs (or composite
bosons\cite{zhk}), which are created {\it in the larger LLL Hilbert
  space} will survive for an extensive set of fractions, i.e., that
one can follow the gapped ground state from weak $V\ll U$ to strong
$V\gg U$ without band crossings. If for $V\gg U$ the CF-bands break up
into groups with degeneracy $1/p$ of an LLL, with distinct groups
separated by energies of order $V$, (with the correct total Chern
index for each group) one can safely use $\bar{\rho}_{LLL}$ in place
of $\bar{\rho}_{T}$ because inter-subband matrix elements of the
former will be suppressed by the large $V$.

For both (i) and (ii), for a given number of flux attached to
transform the electron into a CF, many more fractions are possible
than in the liquid states in the LLL\cite{laugh,zhk,jain}, because the
periodic potential makes it possible to fill a CFLL partially by
creating subbands. The generic problem of the FQHE in a periodic
potential was analyzed by Kol and Read\cite{kol-read} (KR). We use
their analysis to identify the fractional quasiparticle charge and the
Hall conductance for each gapped state. Without Galilean
invariance, $\sigma_{xy}$ need not be the same as the filling
fraction\cite{kol-read}. There is numerical evidence for such states
for lattice hard-core bosons in an external field\cite{moller}.

It will become evident that this approach is easily extended
to two-dimensional time-reversal invariant TIs\cite{TI-reviews,frac-2DTI}.

To illustrate Embedding (i) we consider a periodic square lattice
potential with one flux quantum per unit cell ($a^2=2\pi l^2$). In
order to get nonconstant Chern density ${\cal B}(\bp)$, we need to mix
higher LLs. We will use the basis of states which have the translation
properties appropriate to the potential:

\beq
\psi_{\bp,n}(\bx)=\frac{1}{\sqrt{a}}\sum\limits_{j}e^{ijp_xa+iy(p_y+2\pi j/a)}\Phi_n(x-ja-p_yl^2)
\label{bloch}
\eeq

Here $n$ is a LL index and $\Phi_n$ are  oscillator
eigenfunctions. Now the one-body hamiltonian is

\beq
H=\sum\limits_{\bG}V(-\bG){\hat{\rho}}_e(\bG)=\sum\limits_{\bp}c^{\dagger}_{\bp}h(\bp)c_{\bp}
\eeq where $\bG$ are the reciprocal lattice vectors and $\hat{\rho}_e$
is the electronic density operator in the full Hilbert space including
all LLs, and $h(\bp)$ is a matrix labeled by LL indices.  In computing
the Chern density of the lowest band of $H$, it is important to
remember that the basis functions of Eq. (\ref{bloch}) cannot be
localized\cite{thouless}. Their $\bk$-dependence contributes a uniform
background Chern density of ${\cal B}_{Bloch}=a^2/2\pi$, to be added
to that from the gauge connection of the lowest eigenstate.  One
manipulates the Chern density of the lowest band by tuning the
harmonics of $V$ to approximate the Chern density of the given CB by
the Chern density of this target band, as well as make it flat in
energy. Given this, it is clear that the algebra of
$\bar{\rho}_{T}(\bq)$ will be identical to that of
$\bar{\rho}_{CB}(\bq)$.

To fully exploit the mapping of the CB to the MLLL we express
$\bar{\rho}_{T}(\bq)$ in terms of the LLL projected
density $\bar{\rho}_{LLL}(\bq)$, which satisfies the magnetic
translation algebra\cite{GMP}. By standard techniques\cite{llmix-us}
one can construct a unitary transformation which decouples the target
band from the others. Since the mixing terms between different LLs can
only be superpositions of $\exp{i\bG\cdot\bR^{e}}$ it is evident that

\beq
\bar{\rho}_T(\bq)=\sum\limits_{\bG}r(\bG,\bq)\bar{\rho}_{LLL}(\bq+\bG)
\label{rhot}\eeq

where the function $r(\bG,\bq)$ vanishes as $\bq\to0$ for $\bG\ne0$.
Given the unitary transformation $U_{n_1n_2}(\bp)$ diagonalizing
$h(\bp)$, the coefficients $r$ can be extracted as

\beqr
r(\bG,\bq)=&e^{\frac{(\bq+\bG)^2l^2}{4}+i\frac{l^2}{2}(\bq\times\bG-G_xG_y)}\otimes\nonumber\\&\sum\limits_{\bp,n_1n_2}e^{il^2\bp\times\bG}U^{\dagger}_{0n_1}(\bp')\rho_{n_1n_2}(\bq)U_{n_20}(\bp)
\label{r-coeffs}\eeqr

where $\bp'=(\bp+\bq)mod\bG$, $\rho_{n_1n_2}(\bq)$ are the usual
matrix elements of the density. Generically this $\bar{\rho}_T$ does
not close under commutation\cite{sid}.

To be more specific, consider
$\nu^{CB}=\nu^e=\frac{1}{3}$. Attach two units of flux to the electron
to form CFs, which see $\frac{1}{\nu^{CF}}=\frac{1}{\nu^e}-2=1$
quantum of effective flux per CF, thereby filling 1 CFLL. Since the
target band we have constructed is adiabatically connected to the LLL,
we can expect the Laughlin liquid state of the LLL to be adiabatically connected
to the $\frac{1}{3}$ state in the CB. In particular, the qualitative
properties, such as the ground state degeneracy and Hall conductance,
will be the same as in the corresponding liquid state. Similar logic
holds for $\frac{1}{5}$. These appear to be the states
seen in recent numerics in fractionally filled
CBs\cite{sun,neupert,sheng,bern}.

\begin{figure}
\includegraphics[width=10cm]{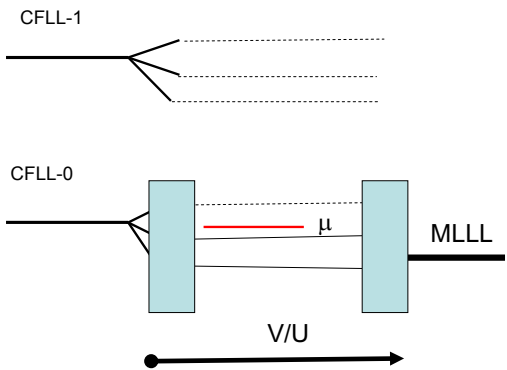}
\caption{A schematic picture of a filling of $\frac{2}{7}$ for
  Embedding (i). The MLLL on the right is the modified lowest band of
  the noninteracting problem with the periodic potential producing
  LL-mixing. The CF's have a filling of $\frac{2}{3}$ and see
  $\frac{3}{7}$ quanta of effective flux per unit cell. Each CFLL
  breaks up into three subbands,and two of the lowest subbands are
  occupied. The grey boxes indicate band-crossing regions where the
  state may not be gapped.
  \label{1phi02/7}}
\end{figure}

However, this does not exhaust the possibilities. Consider
$\nu^{CB}=\nu^e=\frac{2}{7}$. Upon attaching two units of flux, the
CFs see $\frac{1}{\nu^{CF}}=\frac{7}{2}-2=\frac{3}{2}$ quanta of
effective flux per particle, implying $\nu^{CF}=\frac{2}{3}$. In a
translationally invariant system one would expect this state to be
gapless (unless it is paired). However, in a periodic potential the
CFs see $\frac{3}{7}$ quanta of effective flux per unit cell, implying
that each CFLL breaks up into three subbands\cite{TKNN}. One can fill
two of subbands arising from the $n=0$ CFLL and obtain a gapped state,
as shown schematically in Fig. (\ref{1phi02/7}). A surprising property
of this state is that its Hall conductance cannot be $\frac{2}{7}$. To
see this, we use the analysis of KR\cite{kol-read}, who find that
given the total CF-Chern index of all the filled CF-subbands (dubbed
$\sigma_{xy}^{CF}$ in dimensionless units), the quasihole charge and
electronic Hall conductance $\sigma_{xy}$ (again dimensionless) are

\beq
e^*_{qh}=\frac{1}{1+2\sigma_{xy}^{CF}};\ \ \ \sigma_{xy}=e^*_{qh}\sigma_{xy}^{CF}
\label{KR1}
\eeq

and the statistics parameter for the quasiholes is
$\theta=\pi(1-2e^*_{qh})$.

The CF-Chern index of each subband is an integer. For $\frac{2}{7}$
the possibilities for $\sigma_{xy}$ are $\frac{1}{3}$, $\frac{2}{5}$,
$\frac{3}{7}$ etc, depending on precisely what the CF-Chern indices of
the occupied subbands are. While we do not know what $\sigma_{xy}$ is
we do know that it is not $\frac{2}{7}$. This state, not adiabatically
connected to a liquid-like state in the LLL, exists thanks to the
interplay of interactions and periodic potential.

By attaching one quantum of flux, it is possible to map a
hard-core boson problem into a fermion problem (the reverse of the
usual transformation from electrons to Composite Bosons\cite{zhk}). If
the hard-core bosons live in a CB, then one can use the techniques
presented above to analyze their incompressible states with a strong
interaction. In the related problem of hard-core bosons on a lattice
in the presence of external flux, in addition to the standard Laughlin
states\cite{sorensen,moller} there is numerical evidence for states with no
analogue in the continuum\cite{moller}.

\begin{figure}
\includegraphics[width=8cm]{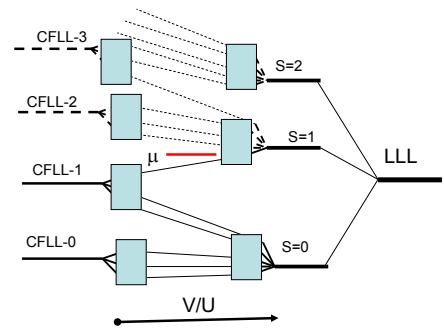}
\caption{An example where $\nu^{CB}={1 \over 5}$ and there are $3$
  flux quanta per unit cell for electrons.  At the far left, we have
  $V=0$ and $U$ is nonzero. There are two filled CFFLL's shown by
  solid lines and some empty CFLL's shown by dotted lines.  At the
  right end, where $U=0$ and $V$ is not, are the three sub-bands of
  the LLL, each having ${ 1\over 3}$ of the LLL states. Since the CF
  sees a flux ${1\over 5}$ as strong as the electron, each CF sub-band
  has ${1 \over 15}$ of the states of the electronic LL. So ultimately
  five of the CF sub-bands must flow into the bottom third of the
  electronic sub-band, their Chern numbers must add to $0$.\label{topobands}}
\end{figure}

We now briefly illustrate Embedding (ii) in Fig. (\ref{topobands}).
Start with 3 electronic flux quanta per unit cell of a periodic
potential. We know \cite{TKNN} that the LLL will split into $p=3$ bands
with a third of the states in each and when the potential contains
only the lowest harmonic only the middle subband will have Chern index
1. This subband is the embedding of the original CB.

Consider $\nu^{CB}={1 \over 5}$. The electronic filling fraction for
(including the filled lower subband) is $\nu^{e}={1 \over 5} \cdot {1
  \over 3}+ {1 \over 3}={2 \over 5}$. Since each electron sees ${5
  \over 2}$ flux quanta, the CFs see half a quantum each and fill
$\nu^*= 2$ CFLLs. Since the CFs see a field that is $B/B^*= \nu^*/\nu
= 5$ times smaller than the electrons, they will see ${3 \over 5}$
quanta per unit cell. This means each CFLL will split into $3$
sub-bands. How many CF-subbands does it take to fill an electronic
subband, each of which has $1/3$ of the states of the LLL? Since the
CF sees a flux ${1\over 5}$ as strong as the electron, each CF
sub-band has ${1 \over 15}$ of the states of the electronic LL. So, if
adiabatic continuity holds, $5$ of the CF sub-bands must flow into
each of the electronic sub-bands as shown in Fig. (\ref{topobands}).

Besides having the right hierarchy of energies, in Embedding (ii) the
subbands merging into the lowest (inert) electronic subband must match its
Chern number, namely $0$, by adiabatic continuity.  This implies that
the single occupied CF-subband going into the target electronic
subband carries a CF-Chern index of 2, which in turn
implies\cite{kol-read} that the Hall conductance of this state is
$2/5$, once again different from the filling of the CB.

Flux attachment offers a way to not only construct excellent
wavefunctions in the LLL\cite{laugh,jain}, but also, via an extended
Hamiltonian approach\cite{rmp} developed by us, a way to compute
dynamical response functions, temperature-dependent properties, and
much more. This approach, which starts from an extended Hilbert space
of CFs supplemented by constraints to ``project'' to the physical
space, has special advantages when applied to our embedding of the CB
into an LLL. In the lattice FQHE problem, while trial wavefunctions
can be constructed, they seem to have a much weaker overlap with the
exact ground state even for Laughlin
fractions\cite{sorensen,moller}. The likely reason is that the density
operator (Eq. (\ref{rhot})), its algebra\cite{sid}, and thus the
entire dynamics, are different. Our approach does not rely on
wavefunctions and is immune to this problem. Secondly, the entire
formalism of ``projecting'' response functions to the physical
subspace (the conserving approximation) carries over unchanged, and
there are no significant finite-size limitations.

In summary, we propose two ways to analyze fractionalized Chern
bands. Embedding (i) uses LL-mixing induced by a periodic
potential $V$ to approximate the Chern density of the CB by that of
the lowest band (the MLLL) in the LL problem by a suitable
manipulation $V$. The density operator of the MLLL (needed for
computations in the interacting theory) can be expressed as a
superposition of LLL-projected density
operators(Eq. (\ref{rhot})). Upon introducing interactions $U$ much
larger than the bandwidth $W$, we argue for the existence of two types
of CF-states with fractional filling in this case: (a) States which
are adiabatically connected to gapped liquid states in the
LLL\cite{laugh,zhk,jain}, which are the gapped states seen in numerics
on CBs\cite{sun,neupert,sheng,bern,sheng-wang} and for FQHE on a
lattice\cite{sorensen,moller}, and (b) More exotic states whose very
existence depends on an interplay of interactions and the lattice
potential. The latter have no analogue in the LLL and a Hall
conductance not equal to the filling factor, and may have been seen in
a problem of lattice hard-core bosons in an external
field\cite{moller}.

Embedding (ii) maps the CB into a subband of an auxiliary LLL split
into electronic subbands by a periodic potential. The density operator
does not have to be projected into the target subband in order to
carry out computations.  If the fractional filling of the CB is such
that the chemical potential lies in a CF-subband gap, and the final
set of CF-subbands break up into groups according to the electronic
hierarchy with the total Chern indices in each group consistent with
the electronic Chern indices, then the dynamics of the fractionally
filled CB is well-approximated by the CFs in a periodic
potential\cite{kol-read}.

Now for 2D time-reversal invariant
TIs\cite{TI-reviews},\cite{frac-2DTI}. We label the pair of Chern
bands making up the noninteracting TI (with equal and opposite Chern
index) by a pseudospin index which need not be conserved and merely
determines the sign of the uniform magnetic fields in the auxiliary
LLL problem. Interactions between CFs with different pseudospins can
be included as before.

We thank the NSF for grants DMR-0703992 and PHY-0970069 (GM),
DMR-0103639 (RS), Yong-Baek Kim, Herb Fertig, Sid Parameswaran, and
Nick Read for illuminating discussions. We are truly grateful to
Shivaji Sondhi and Sid Parameswaran for urging us to confront the
numerics, which led us to Embedding (i). Finally, we are grateful to
the Aspen Center for Physics (NSF 1066293) for its hospitality.

\end{document}